\documentclass{mn2e}
\usepackage{epsfig}
\usepackage{epsf}
\usepackage{graphicx}

\title[Multiple Ejections from RY~Scuti]{Episodic mass loss in binary
  evolution to the Wolf-Rayet phase: Keck and {\it HST} proper motions
  of RY~Scuti's nebula\thanks{Based in part on observations
    made with the NASA/ESA {\it Hubble Space Telescope}, obtained at
    the Space Telescope Science Institute, which is operated by the
    Association of Universities for Research in Astronomy, Inc., under
    NASA contract NAS5-26555.}}

\author[Smith et al.]{Nathan Smith$^1$\thanks{Email:
    nathans@as.arizona.edu}, Robert D.\
  Gehrz$^2$\thanks{gehrz@astro.umn.edu}, Randy Campbell$^3$, Marc
  Kassis$^3$, \newauthor David Le Mignant$^4$, Kawailehua
  Kuluhiwa$^5$, \& Alexei V. Filippenko$^6$
  \\
  $^1$Steward Observatory, 933 N. Cherry Ave., Tucson, AZ 85721, USA \\
  $^2$Astronomy Department, School of Physics and Astronomy,
  University of Minnesota, 116 Church Street SE, Minneapolis, MN 55455, USA \\
  $^3$Keck Observatory, 65-1120 Mamalahoa Hwy, Kamuela, HI 96743, USA \\
  $^4$Laboratoire d'Astrophysique de Marseille, UMR 6110, Univ.\
  Aix-Marseille Provence, 38 rue F.\ Joliot-Curie, F-13388 Marseille, France  \\
  $^5$Department of Physics and Astronomy, University of Hawaii,
  200 West Kawili St., Hilo, HI 96720, USA \\
  $^6$Department of Astronomy, University of California, Berkeley, CA
  94720-3411, USA}

\begin{document}
\date{Accepted 0000, Received 0000, in original form 0000}
\pagerange{\pageref{firstpage}--\pageref{lastpage}} \pubyear{2002}
\def\arcdeg{\degr}
\maketitle
\label{firstpage}

\begin{abstract}
  Binary mass transfer via Roche-lobe overflow (RLOF) is a key channel
  for producing stripped-envelope Wolf-Rayet (WR) stars and may be
  critical to account for Type Ib/c supernova progenitors.  RY Scuti
  is an extremely rare example of a massive binary star caught in this
  brief but important phase.  Its unusual toroidal nebula indicates
  equatorial mass loss during RLOF, while the mass-gaining star is
  apparently embedded in an opaque accretion disk.  RY Scuti's
  toroidal nebula has two components: an inner ionised double-ring
  system, and an outer dust torus that is roughly twice the size of
  the ionised rings.  We present two epochs of $L$-band Keck natural
  guide star adaptive optics (NGS-AO) images of the dust torus, plus
  three epochs of {\it Hubble Space Telescope} ({\it HST}) images of
  the ionised gas rings. Proper motions show that the inner ionised
  rings and the outer dust torus, while having similar geometry,
  came from two separate ejection events roughly 130 and 250 yr
  ago.  This suggests that WR star formation via RLOF in massive
  contact binaries can be accompanied by eruptive and episodic bursts
  of mass loss, reminiscent of luminous blue variables (LBVs).  We
  speculate that the repeating outbursts may arise in the mass gainer
  from instabilities associated with a high accretion rate.  In the
  case of RY Scuti, we know of no historical evidence that either of
  its mass-loss events were observed as luminous outbursts, but if
  discrete mass-loss episodes in other RLOF binaries are accompanied
  by luminous outbursts, they might contribute to the population of
  extragalactic optical transients.  When RLOF ends for RY Scuti, the
  overluminous mass gainer, currently surrounded by an accretion disk,
  will probably become a B[e] supergiant and may outshine the hotter
  stripped-envelope mass-donor star that should die as a Type Ib/c
  supernova.
\end{abstract}

\begin{keywords}
  binaries: eclipsing --- binaries: general --- circumstellar matter
  -- stars: evolution --- stars: mass loss --- supernovae: general
\end{keywords}

\section{INTRODUCTION}

RY Scuti is a remarkable blue supergiant eclipsing binary system at a
distance of 1.8 kpc (Smith et al.\ 2002).  It has a well-determined
period of only 11.1247 days (Smith et al.\ 2002), and the shape of the
eclipse light curve suggests that it is in an advanced stage of
Roche-lobe overflow (RLOF) (Antokhina \& Cherepashchuk 1988; Guircin
\& Mardirossian 1981; Antokhina \& Kumsiashvili 1999; Djurasevic et
al.\ 2001; Melikian et al.\ 2010).  RY Scuti belongs to the class of
W~Serpentis massive binaries, where significant mass transfer has led
to an opaque disk around the mass-gaining star (Plavec 1980), and it
has the shortest known period among examples of the class.

Most recent studies of the system's radial velocity variations
converge on a binary system with an $\sim$8 M$_{\odot}$ primary and a
$\sim$30 M$_{\odot}$ secondary (Antokhina \& Cherepashchuk 1988;
Skul'skii 1992; Sahade et al.\ 1992; Antokhina \& Kumsiashvili 1999;
Grundstrom et al.\ 2007).  The 8 M$_{\odot}$ primary is an O9/B0
supergiant, and is thought to have initially been the more massive of
the two, but has transferred much of its mass to the secondary.  (The
likely initial masses of the primary and secondary were then of order
20-25 and 15-20 M$_{\odot}$, respectively.)  It has been suggested
that the mass-gaining secondary, probably an O5 star, is enshrouded by
an opaque accretion disc (King \& Jameson 1979; Antokhina \&
Cherepashchuk 1988; Antokhina \& Kumsiashvili 1999).  For recent
discussions of the detailed properties of the circumstellar nebula and
the binary system, we refer the reader to Smith et al.\ (2002) and
Grundstrom et al.\ (2007), respectively.  These authors review the
literature concerning spatially resolved structure in the nebula
(Hjellming et al.\ 1973; Gehrz et al.\ 1995, 2001; Smith et al.\ 1999,
2001, 2002), the unusual high-excitation spectrum with multiple-peak
line profiles (Merrill 1928; Swings \& Struve 1940; de Martino et al.\
1992; Skul'skii \& West 1993; Smith et al.\ 2002), and the photometric
and spectroscopic variability of the eclipsing binary (Cowley \&
Hutchings 1976; King \& Jameson 1979; Antokhina \& Cherepashchuk 1988;
Skul'skii 1992; Kumsiashvili et al.\ 2007; Djurasevik et al.\ 2001,
2008; Grundstrom et al.\ 2007).

Detailed study of RY~Scuti and its nebula are of broader interest to
the evolution of massive stars in two chief respects, as follows.

(1) Based on the He-rich abundances of its nebula, the masses of the
stellar components, the orbital configuration, and the evolutionary
state, it has been proposed that the O9/BO supergiant primary will
soon evolve to a Wolf-Rayet (WR) star as a result of binary mass
transfer, and that RY~Scuti therefore represents an immediate
precursor to a massive WR+OB binary system (Guircin \& Mardirossian
1981; Antokhina \& Cherepashchuk 1988; Smith et al.\ 2002).  As such,
it provides a rare glimpse at the formation of WR-like stars in binary
systems, and hence, one of the two chief channels for producing
progenitors of Type Ib/c supernovae (SNe). SNe~Ibc are core-collapse
SNe arising from massive ``stripped-envelope'' progenitors that have
shed their outer H layers, and in some cases their He layers as well;
see Filippenko (1997) for a review.  The first evolutionary channel
for making WR stars is where massive stars with initial masses above
30--35 M$_{\odot}$ shed their H envelopes by virtue of their own mass
loss in stellar winds or eruptions (Conti 1976; Smith \& Owocki 2006).
A second evolutionary channel, which is the only one available to less
massive stars whose winds are too weak to reach the WR phase on their
own, is to have their H envelope (and possibly also the He envelope)
stripped via RLOF in a close binary system (e.g., Paczy\'{n}ski 1967;
Podsiadlowski et al.\ 1992; Petrovic et al.\ 2005).  Recent evidence
from the observed statistics of SNe argues that RLOF may be the
dominant channel for producing progenitors of SNe Ibc (Smith et al.\
2011; see also Yoon et al.\ 2010; Dessart et al.\ 2011).  The
mass-transfer phase in massive binaries is thought to be brief,
lasting only $\sim$10$^4$ yr (Petrovic et al.\ 2005).  RY Scuti is an
extremely rare example of a massive binary star caught in this
critical phase, and it may be the only known example with a bright
spatially resolved circumstellar nebula.\footnote{Another interesting
  example may be the radio-bright source W9 in the Galactic Centre
  region (Dougherty et al.\ 2010), but that source is much farther
  away, its nebula has not been spatially resolved, and it is not an
  eclipsing system.}  Its properties are therefore valuable for
checking the conclusions drawn from studies of WR+OB systems already
in the post-mass-transfer phase.

(2) RY Scuti may provide important clues to formation of toroidal and
bipolar nebulae. Close binaries and mergers are often invoked to
explain the formation of bipolar nebulae and rings like those around
SN~1987A and other massive stars (Morris \& Podsiadlowski 2006;
Collins et al.\ 1999), although asymmetric mass loss from rapidly
rotating stars has also been proposed for such nebulae and disks
(Owocki 2003; Owocki et al.\ 1996; Dwarkadas \& Owocki 2002; Smith
2007; Smith \& Townsend 2007; Chi\c{t}\v{a} et al.\ 2008).  One of the
ambiguities for many nebulae is the lack of independent evidence that
the central stars are (or were) binaries, and the role that binarity
might have played in shaping the nebulae is therefore unknown.
RY~Scuti has the distinct advantage that it is an eclipsing system, so
that its binary stellar parameters are known quite well.  It is in a
state of overcontact where RLOF is occurring and significant mass loss
and mass transfer has taken place.  Its nebula is toroidal, not
bipolar, and so the observed morphology of RY Scuti's mass loss may
provide important constraints on models for shaping nebulae with close
binary influence.

The structure and dynamics of RY Scuti's nebula can aid our
understanding of the role binarity plays in producing SNe Ibc
progenitors and in determining nebular morphology.  The structure,
morphology, and kinematics of the nebula provide clues to its
formation, while the age of its components give a record of the
system's recent mass-loss history.

Previous observations by Gehrz et al.\ (1995, 2001) and Smith et al.\
(1999, 2001, 2002) have established that RY Scuti is $1.8 \pm 0.1$ kpc
distant and that its toroidal nebula is separated into two components:
an outer dust torus with a diameter of $\sim$2\arcsec\ or 3600 AU, and
an inner ionised torus with a diameter of $\sim$1\arcsec\ (1800 AU).
The mass of the dust torus is $\sim$10$^{-6}$ M$_{\odot}$ (dust only),
and the ionised inner component has a gas mass of at least 0.003
M$_{\odot}$.  The ionised component has an unusually high-excitation
spectrum (Merrill 1928; Swings \& Struve 1940; de Martino et al.\
1992; Smith et al.\ 2002), and displays evidence for significant He
and N enrichment (Smith et al.\ 2002).  In high-resolution images
taken with the {\it Hubble Space Telescope} ({\it HST}) by Smith et
al.\ (1999, 2001), the inner ionised torus appears to break up into a
pair of plane-parallel rings, analogous to the polar rings of
SN~1987A, but confined much more closely to the equatorial plane
(i.e., at latitudes of $\pm$14\arcdeg\ from the equator, rather than
$\sim$45\arcdeg\ as in SN~1987A).  The ionised rings are expanding
with Doppler shifts of roughly $\pm$42 km s$^{-1}$, and an initial
(although imprecise) measurement of their proper-motion expansion has
been made in two epochs of {\it HST} images separated by $\sim$2 yr,
combined with two epochs of radio continuum images obtained with the
Very Large Array that were separated by 9 yr. These data implied an
ejection episode sometime in the late 19th century (Smith et al.\
2001).  The kinematics of the outer dust torus were unknown before the
present study.

In this paper, we present a third epoch of {\it HST} images, extending
the time baseline for proper motions made with the same instrument to
more than a decade.  We also present the first adaptive-optics (AO)
images of RY Scuti obtained in the thermal infrared (IR), providing
the sharpest picture yet of the structure in the outer dust torus.  We
obtained two epochs of AO images in the same filter with the same
instrument, separated in time by 6 years, and we use these to measure
for the first time the expansion rate of the dust torus separately from 
the ionised gas. We describe the observations in \S 2, the
multi-wavelength morphology in \S 3, and the
results from proper-motion measurements in \S 4.  In \S 5 we discuss
implications for the formation of non-spherical nebulae and for SN~Ib/c 
progenitors, and speculate about optical transients associated
with episodic RLOF events in massive binaries.  We summarise our
conclusions in \S 6.

\begin{table}\begin{center}\begin{minipage}{3.3in}
      \caption{Imaging Observations of RY Scuti}
\scriptsize
\begin{tabular}{@{}llcl}\hline\hline
Date        &Tel./Instr.      &Filter  &Exp.  \\ \hline
1997 Jun 01 &{\it HST}/WFPC2  &F656N   & 5s, 2$\times$20 s, 2$\times$120 s  \\
2000 Feb 21 &{\it HST}/WFPC2  &F656N   & 2$\times$10 s, 2$\times$120 s \\
2009 Apr 19 &{\it HST}/WFPC2  &F656N   & 2$\times$10 s, 2$\times$260 s \\
2009 Apr 19 &{\it HST}/WFPC2  &F658N   & 2$\times$18 s, 2$\times$350 s \\
2003 Jun 11 &Keck/NIRC2 AO    &$L_p$   & 5$\times$10.6 s, 53 s total   \\
2009 Aug 27 &Keck/NIRC2 AO    &$L_p$   & 10$\times$5.3 s, 53 s total   \\
\hline
\end{tabular}\label{tab:letters}
\end{minipage}\end{center}
\end{table}

\section{OBSERVATIONS}

We obtained multi-epoch high-resolution observations of RY Scuti, using 
{\it HST} images to trace the inner ionised rings and Keck Observatory
$L$-band images to trace the outer dust torus in the IR.  The log of
observations is listed in Table 1.

\begin{figure*}\begin{center}
    \includegraphics[width=5.1in]{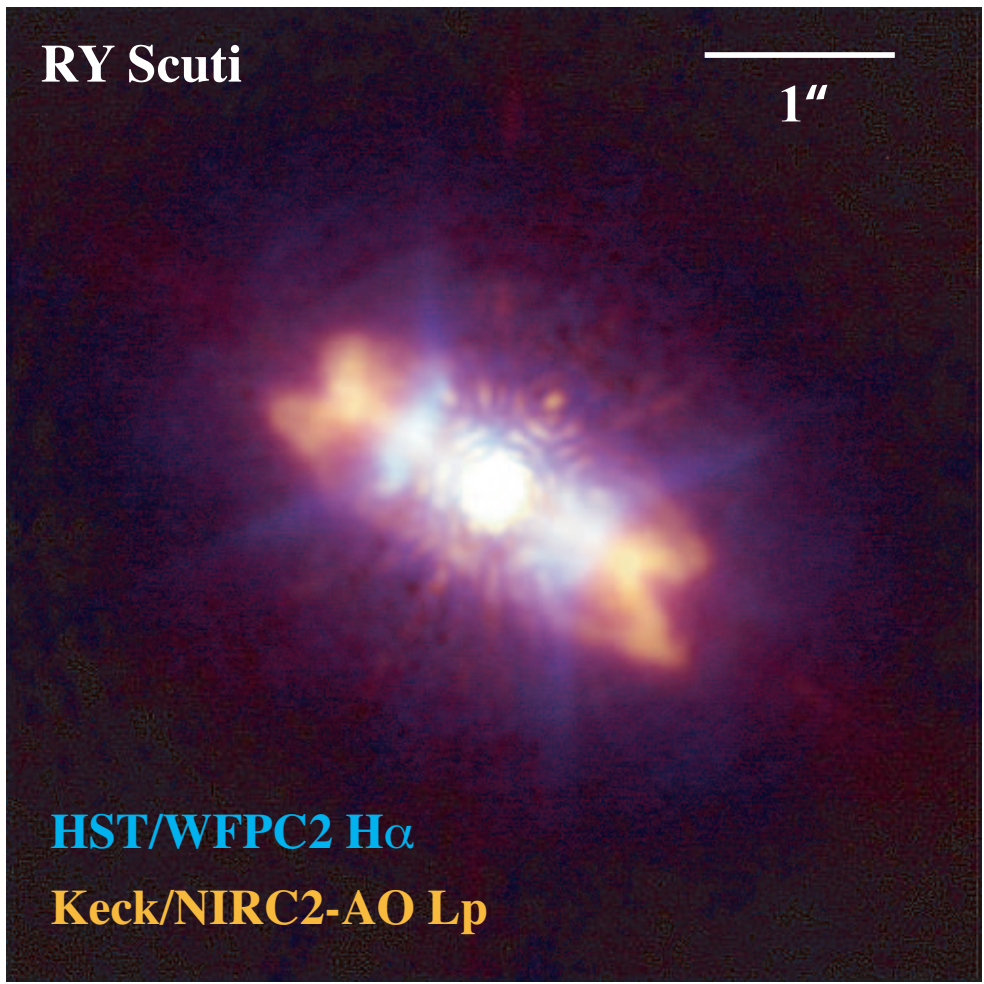}
\end{center}
\caption{A colour composite of the ionised gas traced by the {\it HST}/WFPC2
  image taken through the F656N H$\alpha$ filter (blue-green) and the
  warm dust traced by the Keck/NIRC2-AO image in the $L_p$ filter
  (red-orange).}
\label{fig:color}
\end{figure*}

\subsection{Multi-Epoch HST Imaging}

Three epochs of images of RY Scuti were obtained with the {\it HST} Wide
Field Planetary Camera 2 (WFPC2) using the F656N (H$\alpha$) filter,
plus single epochs with the F658N ([N~{\sc ii}] $\lambda$6583; see
below) and F953N ([S~{\sc iii}] $\lambda$9532) filters.  The first two
epochs of H$\alpha$ images were published and analyzed previously
(Smith et al.\ 1999, 2001).  For the third epoch of H$\alpha$
observations, we implemented the same observing strategy and followed
the same data-reduction steps as for the first two.  This involved
combining a series of exposures with a range of exposure times to
correct for CCD blooming from the bright central star, as well as a
careful subtraction of a model point-spread function (PSF) generated
by the TinyTim software, as described in the earlier papers.  The PSF
subtraction is necessary because the extended diffraction pattern in
the PSF from the bright central star can interfere with and mask the
nebular structures.  After PSF subtraction, the newest epoch of images
confirms the same structures seen in our earlier studies, but taken
with a different roll angle and orientation of the PSF diffraction
spikes.

Our goals in obtaining a third epoch were to confirm our previous
detection of expansion of the rings made with a short temporal baseline
(Smith et al.\ 2001) and to thereby improve the precision of the
ejection age for comparison with that of the outer torus.  For the
proper-motion measurements, we use the F656N (H$\alpha$) filter as it
is the only one available in all three epochs.  The F658N image is
useful to investigate any possible spatial gradients in ionization
structure in the rings.  The elapsed time between the second epoch and
the first is 994.43 days, or $\Delta t$ = 2.723 yr.  The last epoch
extends this temporal baseline to 4340.13 days, or $\Delta t$ = 11.883 yr
since epoch 1.

\subsection{Keck NIRC2-AO Imaging}

Using the Keck II AO system (Wizinowich 1999) with the instrument
NIRC2, we obtained two epochs of $L$-prime (hereafter $L_p$) images. 
They were acquired with the narrow camera (10 mas
pixel$^{-1}$ scale) making use of a five-point box dither pattern
with a 1$\farcs$0 step size. Due to the peak brightness of the RY~Scuti
central point source, a square subarray of 512 pixels (out of 1024)
was employed so that the minimum exposure time could be set to 50 ms,
thus avoiding saturation. The dominant source of background at 3.8
$\mu$m is the thermal radiation of the AO system telescope
optics. Dust on the 11 ambient-temperature optical surfaces that are
in the path of NIRC2 are a significant source of background. Frequent
dithers were used to reduce time varying effects of the
background. RY~Scuti ($V = 9.14$ mag) provides plenty of visible-light
flux to the AO wavefront sensor, allowing the frame rate to be set to
a relatively high frequency; it was 700 Hz in 2003 and
1500 Hz in 2009. Higher frequencies became possible thanks to an
upgrade of the wavefront controller (van Dam 2007).  The AO
performance was superb, in both cases resulting in diffraction-limited
resolution of about 72 mas with a Strehl greater than 70\%. The PSF is
well sampled in the NIRC2 narrow-camera plate scale.

The data reduction was performed in the customary method for IR
imaging data. Before shifting and coadding the 3-point dithered data,
we flat-fielded the images using normalised flats constructed from
sky-only data acquired 1$\arcmin$ north of the nebula.  The flattened
images were then background subtracted using the sky images. The
images were aligned using the centroid of RY~Scuti's central point
source. The initial goal was to produce the best image yet of the IR
dust torus with diffraction-limited imaging in the near-IR to
complement the previous Keck mid-IR images obtained with the 
long-wavelength spectrometer (Gehrz et al. 2001).  
The subsequent goal of the later-epoch observation was to
measure the expansion of the outer dust torus separately from the
inner ionised gas nebula.

\begin{figure*}\begin{center}
\includegraphics[width=6.3in]{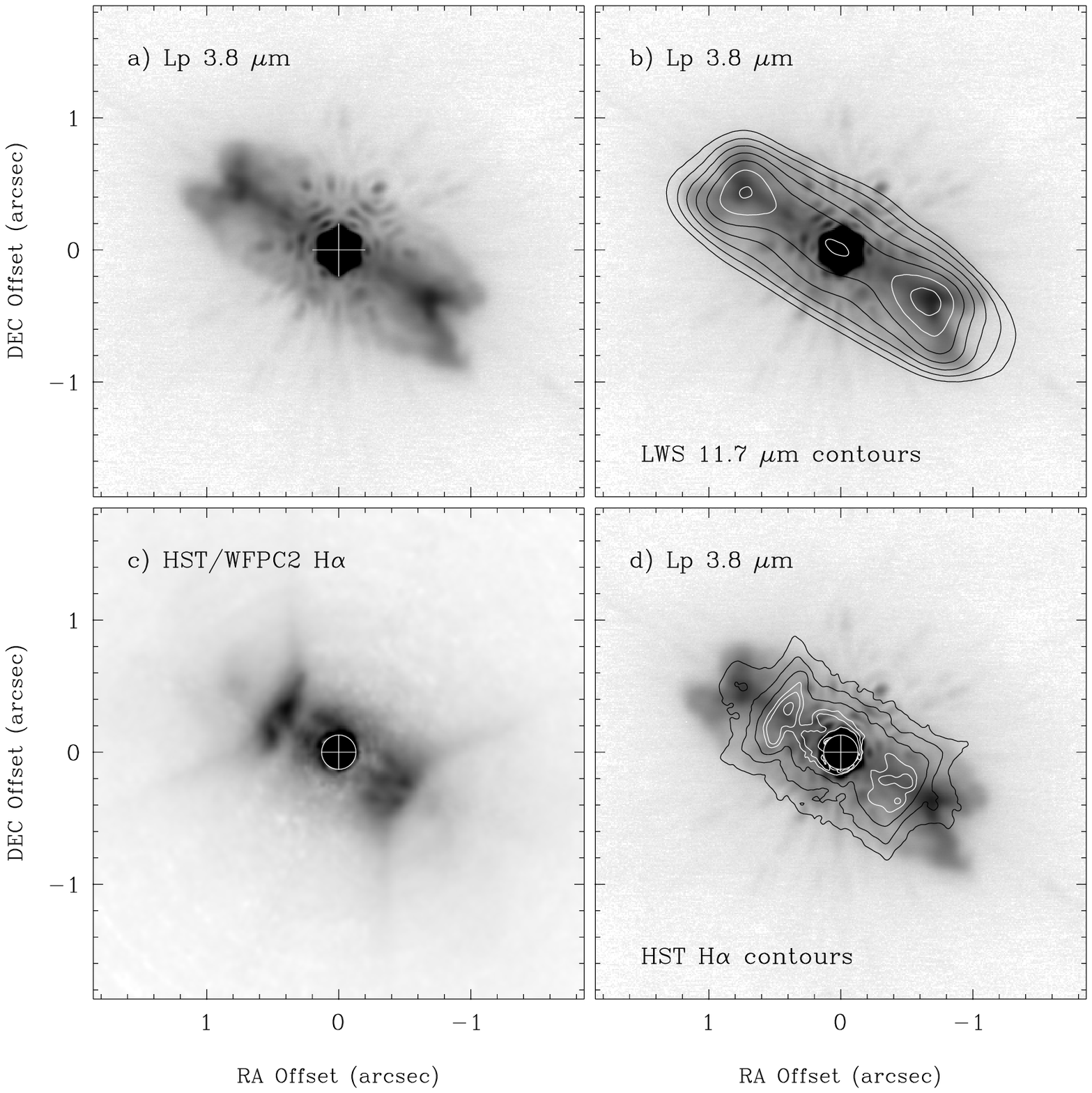}
\end{center}
\caption{(a) The first-epoch 2003 June 11 $L_p$-band NIRC2-AO image of
  RY Scuti (the second epoch looks identical, except for a different
  orientation of the diffraction pattern).  (b) Same image as in
  (a), but overlaid with contours of the 11.7 $\micron$ image from
  Gehrz et al.\ (2001).  (c) The PSF-subtracted {\it HST}/WFPC2
  H$\alpha$ (F656N) image of RY Scuti.  (d) The $L_p$-band image from
  (a) overlaid with contours of the {\it HST} image.}
\label{fig:contour}
\end{figure*}

\section{MULTI-WAVELENGTH STRUCTURE}

We investigated the multi-wavelength structure of the nebula by
comparing the IR and optical images.  The morphology of the optical
images in H$\alpha$ was already discussed in detail in previous papers
(Smith et al.\ 2002, 2001, 1999).  The NIRC2-AO images in the $L$ band
provide the highest-resolution images of the dust torus so far.  They
are comparable in spatial resolution to the {\it HST} images,
permitting the first meaningful comparison between the two.
Figure~\ref{fig:color} shows a colour composite comparing the {\it HST}
image to the near-IR Keck-AO $L_p$ image, while
Figure~\ref{fig:contour} provides various comparisons of multi-wavelength
data from the present study and from a previous study of thermal-IR
Keck images by Gehrz et al.\ (2001).

The first-epoch 2003 image of RY Scuti obtained with NIRC2-AO is shown
in Figure~\ref{fig:contour}a.  The combination of AO and the better
diffraction limit at $\sim$4 $\mu$m as compared to $\sim$10 $\mu$m
means that this $L_p$-band image provides the sharpest view yet of the
structure in the dust torus around RY Scuti, improving the spatial
resolution from our previous thermal-IR Keck images without AO (Gehrz
et al.\ 2001) by a factor of 2--3.  The warm-dust emission in this
filter traces the same dust responsible for the torus observed at
longer wavelengths.  This is evident from Figure~\ref{fig:contour}b,
which shows the same NIRC2-AO image from Figure~\ref{fig:contour}a 
with the contours of 11.7 $\mu$m emission from Gehrz et al.\ (2001) 
superposed. Allowing for differences in spatial resolution (and for stronger
photospheric emission from the central star at shorter wavelengths),
the $L_p$-band image has the same spatial distribution as the 11.7
$\mu$m thermal-IR emission.  This was expected, since the 2--20 $\mu$m
spectral energy distribution (SED) of the dust torus can be fitted
with a single dust temperature (Gehrz et al.\ 2001).  In other words,
the $L_p$ filter at 3.8 $\mu$m samples the Wien tail of the same
300--400 K dust whose emission peaks at $\sim$10 $\mu$m, rather than
sampling hotter dust closer to the star.  The new AO image indicates a
very thin distribution of dust in the radial direction, consistent
with our findings below that the dust torus originated in an episodic
ejection from the star.

\begin{figure*}\begin{center}
\includegraphics[width=5.3in]{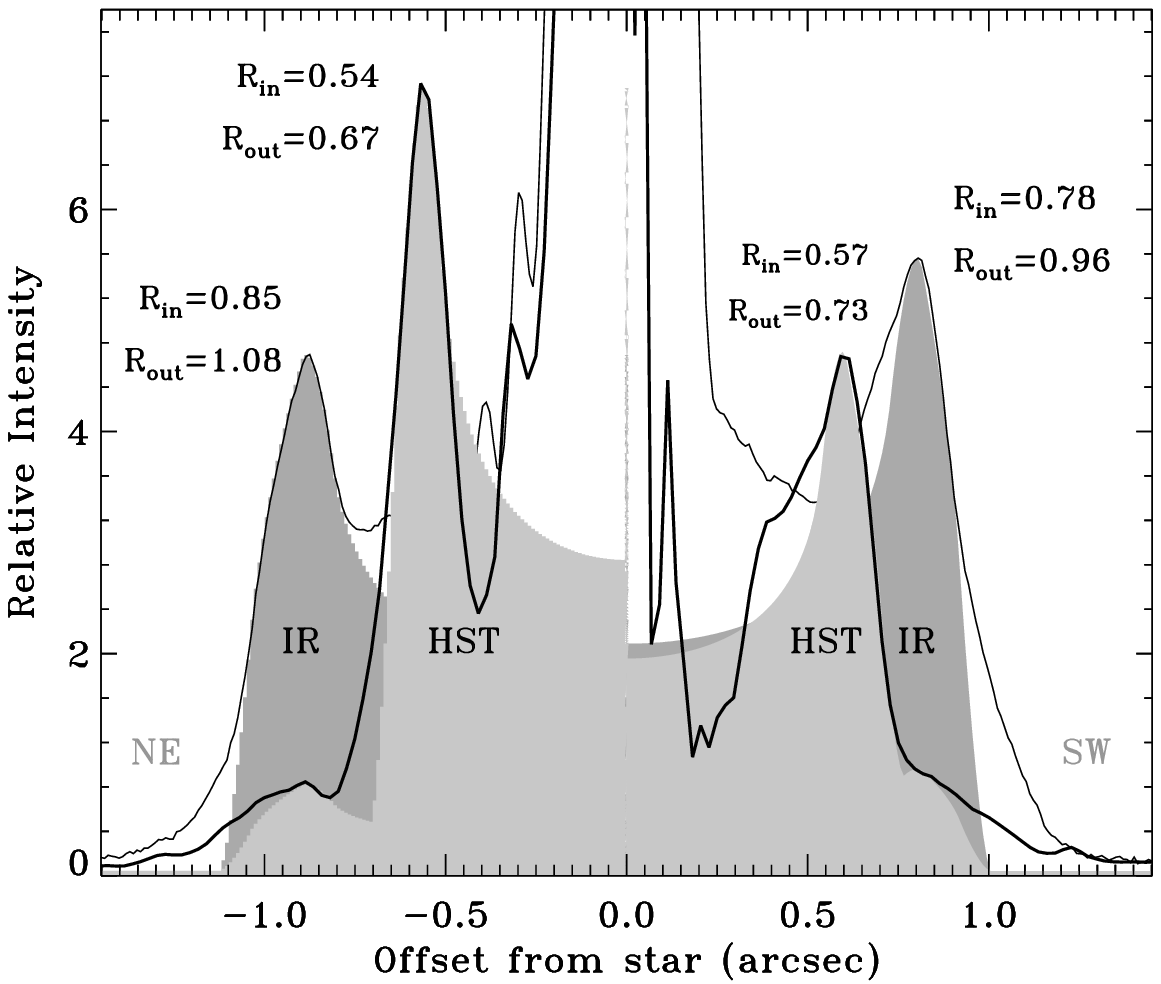}
\end{center}
\caption{Spatial intensity scans across the major axis in the 2009
  PSF-subtracted {\it HST} H$\alpha$ image (thick line) and the 2009
  Keck IR image (thin line), along a position angle of 60\arcdeg.  For comparison,
  the shaded regions show simple models for the intensity of a slice
  through an idealised optically thin edge-on torus, with inner and
  outer radii noted above each peak.  We used different parameters for
  the NE and SW sides due to the inherent asymmetry of the nebula.
  The darker and lighter shaded regions respectively approximate the 
  intensity cut across the IR torus and the inner ionised rings seen 
  by {\it HST}.  This lighter shaded region
  includes a broader base that is an arbitrarily scaled version of the
  IR torus (to mimic the scattered H$\alpha$ light from the outer
  nebula; see Smith et al.\ 2002).}
\label{fig:model}
\end{figure*}

The NIRC2-AO images show unprecedented detail of the structure in the
IR torus.  The toroidal nebula obviously appears pinched at the waist,
flaring above and below the equator.  This appearance could be due to
the overlap of two inclined rings, as in the optical images, or it
could be that the dust torus is akin to an hourglass structure with
the top and bottom chopped off.  In either case, the dust torus
appears to have a sharp outer boundary, and we see no evidence in
either the {\it HST} or near-IR images for faint extensions of a
larger hourglass structure.  Overall, the structure of the IR dust
torus resembles the morphology seen in the inner ionised rings, but
with roughly twice the size.

The near-IR AO images reveal no clear evidence for an enhancement of
dust emission from parts of the nebula inside the dust torus,
consistent with the single-temperature dust SED as noted above.  In
particular, there is no enhancement of IR emission coincident with the
ionised structures of the inner rings seen in H$\alpha$ images with
{\it HST} (except for some spurious features associated with the Keck
PSF).  Since the inner rings are closer to the star and any dust
therein would be hotter and more easily detected at short IR
wavelengths, we conclude that the inner rings have a much lower dust
mass than the outer rings.  The second mass-ejection event that
produced the inner rings evidently did not form dust as efficiently --
at least not yet.  Given that the temperature of the outer dust torus
is 300--400 K (Gehrz et al.\ 2001), the equilibrium grain temperature
in the inner rings (at roughly half the distance from the same star)
should be about 420--560 K.  Since the condensation temperature of
dust grains is typically $\ga$1000 K, any dust that was destined to
form in the inner rings should have done so already.  An interesting
possibility is that the second ejection, which formed the inner
ionised nebula, was able to shield the outer torus from ionising
radiation, thereby allowing it to form dust.  One could speculate,
then, that a hypothetical future ejection might be needed to shield
the inner torus seen now in {\it HST} images in order to facilitate
dust formation in that feature.

The higher resolution of the new near-IR AO images reveals for the 
first time a clear gap between the inner ionised rings and the outer dust 
torus. Figure~\ref{fig:model} shows tracings across the middle of the
Keck IR and {\it HST} images (see below), both as observed in 2009.
For comparison, Figure~\ref{fig:model} also shows very simple,
idealised models for the optically thin intensity from a cross section
through the middle of the geometrically thin shell (see Smith et al.\
2007 for more details). These simple models are not perfect matches to
the data due to the non-azimuthally symmetric density structure of the
ionised rings (Smith et al.\ 2002).  While the NE side of the nebula
matches the simple models rather well, the SW portion deviates --- and
interestingly, both the inner ionised rings and the outer dust torus
deviate from the model on the SW side in the same manner.  This
requires that both the inner and outer tori {\it deviate from
  azimuthal symmetry in a similar way}.  This is probably an important
clue to the ejection mechanism that formed them.

The simple models in Figure~\ref{fig:model} allow us to provide rough
estimates of the inner and outer radii for both the ionised and dusty
tori.  These values are listed in Figure~\ref{fig:model}, and we note
that the values derived from images of the inner ionised component
agree with the thickness inferred from STIS spectroscopy (Smith et
al.\ 2002).  The radial thicknesses of the model shells are
23--28\% of their radii.  More importantly, there is a clear gap
between the inner edge of the IR torus and the outer edge of the
ionised rings (i.e., the structures do not touch or overlap).  This is
most evident on the NE side of the nebula (Figure~\ref{fig:model}),
and is clear in the colour image in Figure~\ref{fig:color} as well,
indicating a spatial separation of the dust and ionised gas.
Physically, this is quite meaningful; it indicates that the outer
boundary of the ionised rings is not caused by a simple ionization
front, because in that case we would expect non-ionising UV photons to
penetrate the ionization front and heat the dust immediately outside
it.  The large separation instead suggests that the IR peak is a
separate density enhancement at a larger radius where the remaining UV
radiation is absorbed. (The inner ring may nevertheless play an
important role in shielding the outer dust torus, but our point here
is that the outer boundary of the ionised torus is caused by a drop in
density, not an ionization front.)  As we show below, proper motions
indicate that the outer IR dust torus is older and originated in a
separate mass-ejection event.  The spatial gap between the inner and
outer rings indicates that the outer ring was not responsible for
shaping the inner ring, because they are not interacting
hydrodynamically.

In the ionised inner torus, the NE peak is much brighter than the SW
peak.  For the IR torus, the reverse is true.  Perhaps this is because
the NE side of the IR torus is shielded by a denser inner nebula, as
compared to the SW side.  Of course, these differences may be due
purely to different density distributions as well.

\begin{figure*}\begin{center}
\includegraphics[width=4.9in]{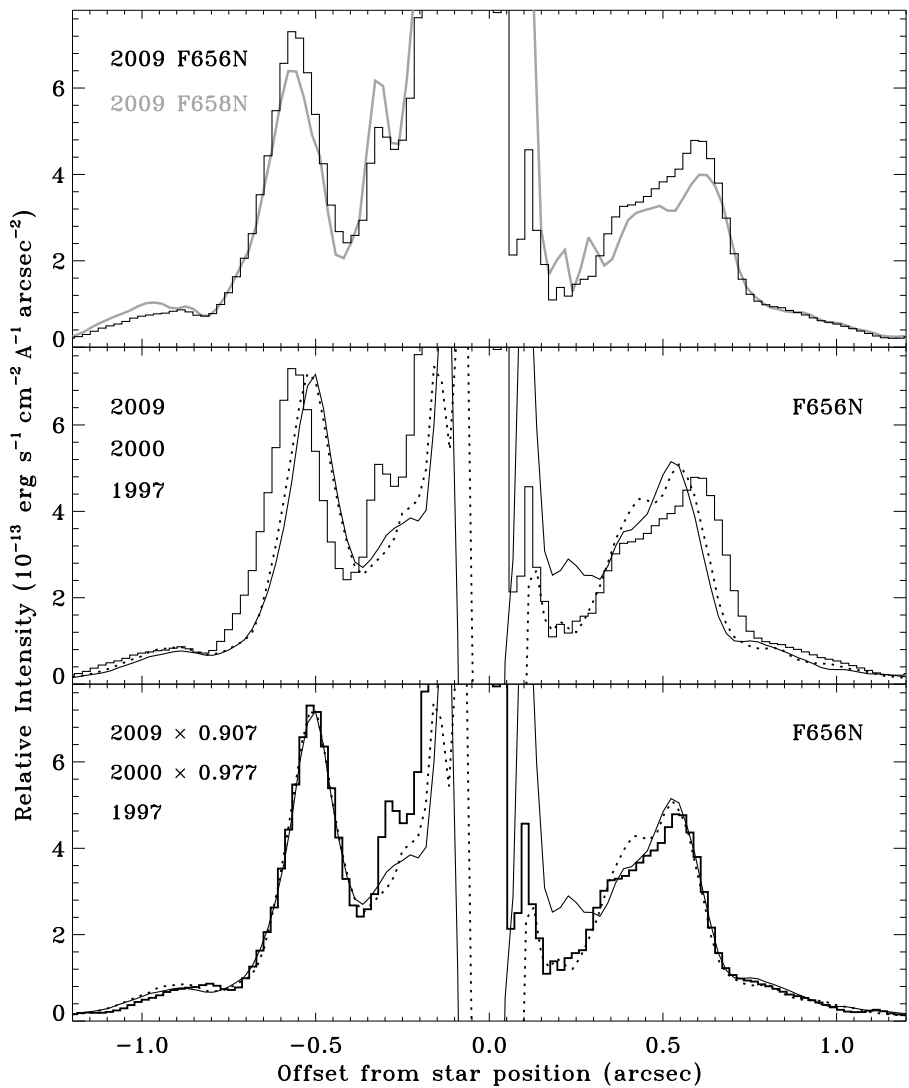}
\end{center}
\caption{Spatial intensity scans across the major axis in
  PSF-subtracted {\it HST} images.  The top panel shows tracings for
  H$\alpha$ F656N (solid black) and [N~{\sc ii}] F658N (solid grey) in
  the newest epoch from 2009.  The middle panel gives the tracings
  through H$\alpha$ images at all three epochs in 1997 (solid), 2000
  (dotted), and 2009 (solid histogram).  The bottom panel is the same
  as the middle, except that the sizes of the 2000 and 2009 intensity
  scans have been multiplied by 97.7\% and 90.7\% of their observed
  size, respectively, to match the size in the 1997 image.}
\label{fig:hstpm}
\end{figure*}

\begin{figure*}\begin{center}
\includegraphics[width=4.9in]{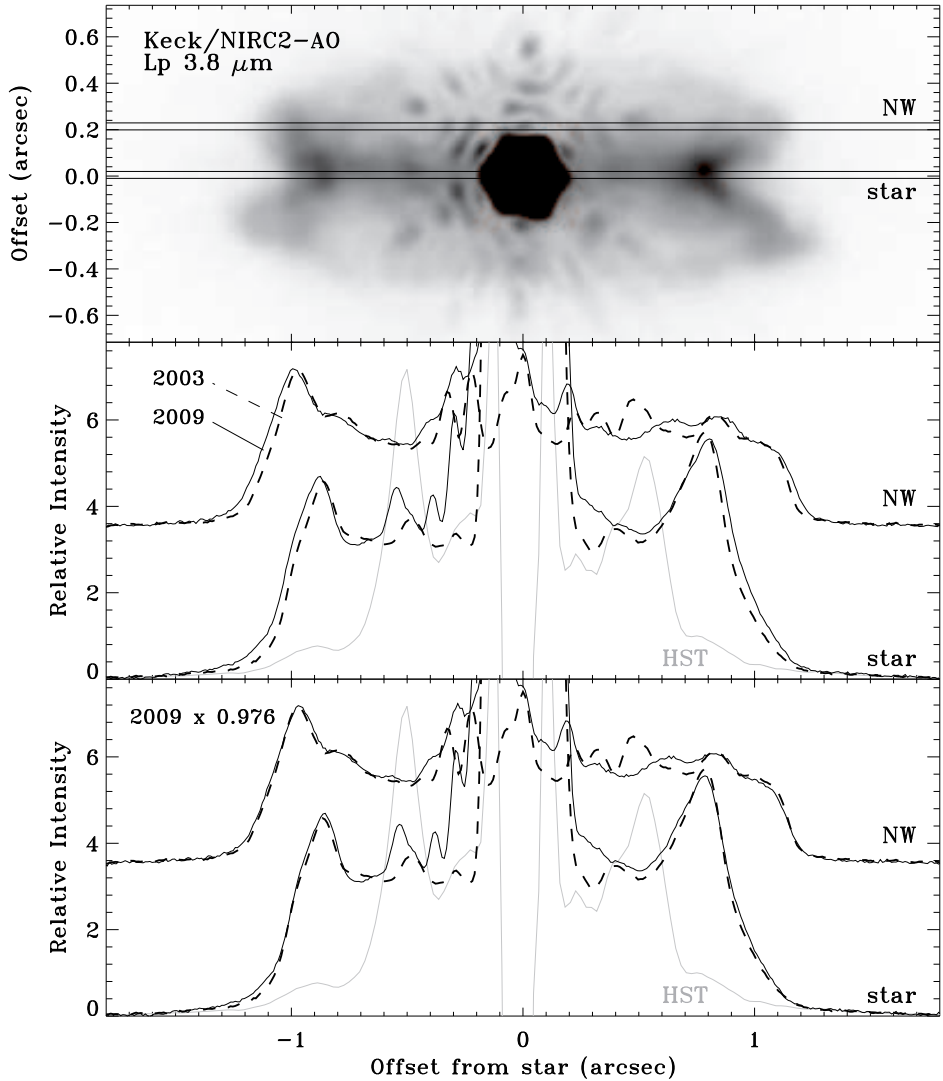}
\end{center}
\caption{Proper motions of the IR torus in Keck/NIRC2-AO images.  The
  top panel shows the 2003 June 11 image of RY~Scuti in the $L_p$ band
  obtained with Keck/NIRC2-AO, rotated counterclockwise by 30\arcdeg\
  (note that many features within 0$\farcs$5 of the star are part of
  the PSF diffraction pattern of the Keck telescope).  The horizontal
  lines marked ``star'' and ``NW'' give the positions of
  intensity tracings across the image, shown in the middle and bottom
  panels.  The middle panel displays the intensity scans for these two
  positions (with a constant offset added to the ``NW'' tracing) as
  observed in 2003 June (dashed) and 2009 August (solid).  The bottom
  panel is the same as the middle, except that the size of the 2009
  intensity scan has been multiplied by 97.6\%, giving the best match
  to the 2003 image (both ``star'' and ``NW'' tracings were scaled by
  the same factor).  The light-grey curve shows the same tracing of
  H$\alpha$ from the {\it HST} data in Figure~\ref{fig:hstpm}.}
\label{fig:keckpm}
\end{figure*}

\section{PROPER MOTIONS OF THE TWO COMPONENTS}

\subsection{The Inner Ionised Rings with HST}

By aligning the three epochs of {\it HST} images, we were able to
confirm that the expansion of the rings is primarily homologous
(i.e., self-similar radial expansion).  When an image taken at an early
epoch is magnified by some scaling factor, it looks identical to one
taken at a later epoch.  We detect no evidence for non-radial motion,
acceleration, or deceleration in the expanding nebula.

To quantify the fractional increase in size between epochs, we scaled
later images so that the nebular structure matched that in the first
epoch, and then performed a cross correlation of the scaled image to
estimate the best scaling factor.  This is similar to the method
employed by Morse et al.\ (2001) for measuring proper motions in {\it
  HST} images of $\eta$ Carinae, except that here we adopted a
multiplicative size-scaling factor for the whole nebula (with central
regions near the star masked out) rather than measuring translational
shifts independently for many individual small condensations.  The
former method is well suited to the compact size and simpler structure
of RY Scuti's nebula.

To illustrate the scaling between epochs, Figure~\ref{fig:hstpm}
(middle panel) shows intensity tracings across the major axis of the
nebula through the emission peaks on either side of the star at each
epoch.  One can see that the nebula is clearly expanding with time.
The bottom panel in Figure~\ref{fig:hstpm} shows the same tracings,
but with the pixel size scale of the 2009 and 2001 images reduced to
match the structure in the 1997 image.  Similar tracings for the Keck
images are shown in Figure~\ref{fig:keckpm}.


We define the scale factor $\epsilon$ by which the size of the nebula
at epoch 2 ($R_2$) must be reduced to match the radius at epoch 1
($R_1$) as $\epsilon$ = $R_1$/$R_2$.  In terms of this scaling factor
$\epsilon$, the age of the nebula $t$ (the time since ejection
relative to epoch 2) is given by

\begin{equation}
  t = \Delta t \ \times \ (1 - \epsilon)^{-1}, 
\end{equation}

\noindent 
where $\Delta t$ is the time period that has elapsed between
the images at epochs 1 and 2.  We define epochs 1, 2, and 3 as the
images taken in 1997, 2000, and 2009, respectively.  From the dates of
observations, $\Delta t_{1,2} = 2.7226$ yr and
$\Delta t_{1,3} = 11.8826$ yr.  By cross correlating the NE and SW peak
structures at different epochs, we measure
$\epsilon_{1,2} = 0.977 \pm 0.003$ and $\epsilon_{1,3} = 0.907 \pm 0.003$.
Equation (1) then yields ages of $t_{1,2} = 118.4 \pm 16$ yr (relative
to 2000) and $t_{1,3} = 127.8 \pm 4.2$ yr (relative to 2009).  The
uncertainties are Gaussian 3$\sigma$ error bars resulting from the
spatial cross-correlation of each pair of epochs.  Thus, the baselines
between epochs 1 and 2 (1997--2000) and between epochs 1 and 3
(1997--2009) both agree on an ejection date around the year 1881 for
the ionised torus.  We adopt an uncertainty of $\pm$4.2 yr in the age
and ejection date, since it corresponds to the longer temporal baseline
with a more precise measurement.  Note that this measurement is
independent of the precision to which we can spatially align the
central star in the images, since we are simply measuring the growth
in size across the nebula, not the distance from the star.

\begin{figure}\begin{center}
\includegraphics[width=3.1in]{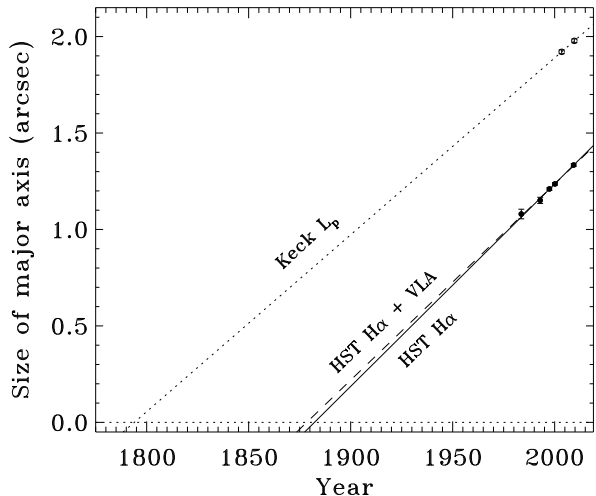}
\end{center}
\caption{Plot of the increase in the width of the major axis of RY
  Scuti's nebula with time; the width is measured consistently at half
  the peak intensity level.  Unfilled points correspond to the IR
  torus measured in Keck $L_p$ images obtained in 2003 and 2009, while
  the filled points correspond to the ionised rings measured in
  archival VLA data from 1983 and 1992 (from Smith et al.\ 2001) and
  {\it HST}/H$\alpha$ images obtained in 1997, 2000, and 2009.  The
  dotted line extrapolates the expansion rate measured from the two
  Keck points.  The dashed and solid lines show least-squares fits 
  for the expansion of the ionised rings with the first two VLA data 
  points included and excluded, respectively, along with the 
  subsequent three {\it HST} images. }
\label{fig:pmplot}
\end{figure}


Figure~\ref{fig:pmplot} shows the expansion of RY Scuti's nebula in a
different way.  This is an updated version of Figure~3 from
Smith et al.\ (2001), including the new Keck data and the last epoch
of {\it HST} imaging.  It shows measurements of the width of the major
axis of the nebula measured at half the peak intensity.  For the
ionised rings, this includes the images of radio free-free emission
obtained with the VLA in 1983 and 1992, as well as the three epochs of
H$\alpha$ imaging in 1997, 2000, and 2009.  The dashed and solid lines
show a linear least-squares fit (weight $\propto$ $\sigma^{-1/2}$) to
the expansion rate of the ring with (dashed) and without (solid) the
early VLA data included in the fit.  These fits yield ejection dates
of 1878$\pm$5 and 1883$\pm$6, respectively, for the ionised component
of the nebula, consistent with the ejection date of 1881$\pm$4 derived
from the first method discussed above.  The dotted line extrapolates
the expansion rate measured the same way in the IR Keck images; this
is not a fit since there are only two points.  The implied ejection
date of 1794$\pm$30 yr for the dust torus is younger than that derived
below, but is consistent within the uncertainty.

\subsection{The Outer Dust Torus with Keck AO}

We employed the same method of measuring the expansion of the dust
torus in IR images as described above for the {\it HST} images, except
that for the IR images we had only two epochs.
Figure~\ref{fig:keckpm} shows intensity tracings across the long axis
of the dust torus at two positions, indicated in the top panel.  The
middle panel displays the tracings for the two epochs at the two sampled
positions, while the bottom panel illustrates the same tracings except that
the 2009 epoch has been scaled to match the 2003 image (see discussion
above concerning the {\it HST} images and Fig.~\ref{fig:hstpm}). The
two different position samples are consistent with the same expansion
rate.

The comparison of the 2003 and 2009 Keck images of the IR dust torus
shows that the size of the torus in 2003 was 97.6\% ($\pm$0.3\%) of the
size observed in 2009 (i.e., $\epsilon_{1,2} = 0.976 \pm 0.003$).  Since
the time elapsed between images was $\Delta t = 6.216$ yr, Equation
(1) indicates that the dynamical age of the dust torus is $255 \pm 32$
yr, implying an ejection date around the year 1754.  This is more than
twice the age of the inner ionised rings determined above from 
{\it HST} images, so the two components cannot be the result of the same
mass ejection.  The age derived from this method is consistent within
the uncertainty with the age derived from the simpler method described
above and shown in Figure~\ref{fig:pmplot}, where we measured
the changing width of the major axis of the nebula at half the peak
intensity, following Smith et al.\ (2001).  This simpler method is
more susceptible to error caused by changes in spatial resolution,
which is more of a concern in ground-based data than in {\it HST} data
having a more consistent PSF.  We therefore favour the ejection date
around 1754.

\section{DISCUSSION}

\subsection{Formation of the Double Rings}

The mechanism for the formation of the rings around SN~1987A has
presented an enduring mystery that has no obvious answer (e.g.,
Blondin \& Lundqvist 1993; Burrows et al.\ 1995; Martin \& Arnett
1995; Collins et al.\ 1999; Morris \& Podsiadlowski 2005), but one
expects that the formation mechanism of the rings might also be an
important clue to understanding the peculiar nature of the blue
progenitor star Sk $-$69~202.  Smith et al.\ (2007) noted a population
of stars with equatorial rings that are analogous to the equatorial
ring around SN~1987A, including RY Scuti.  Only one other object,
the luminous blue variable (LBV) star HD~168625, is known to have
triple rings similar to those of SN~1987A (Smith 2007).  RY Scuti has a 
set of double ionised rings which are not identical to those of SN~1987A,
but may be related.

One can test some models for formation of double rings by measuring
the expansion dynamics of the rings -- i.e., do they exhibit
homologous expansion, non-radial expansions, or other peculiar motions
or time-variable illumination?  Models with aspherical ejection from
the star system would predict homologous expansion at these large
distances.  On the other hand, models such as those of Chi\c{t}\v{a}
et al.\ (2008) have the rings produced by a bipolar wind pushing
through a previously ejected thin shell, with the two structures
colliding and forming rings at their intersection.  In this model, the
apparent ring structure originates where the rings are observed,
rather than the shaping mechanism arising close to the star.  It would
predict motion of the rings toward the equatorial plane with time as the
inner bipolar wind sweeps through the thin spherical shell, with the
intersection migrating from the pole to the equator (see Chi\c{t}\v{a}
et al.\ 2008).  We therefore conclude that models such as the one
discussed by Chi\c{t}\v{a} et al.\ (2008) do not apply in the case of
RY Scuti, since the nebula is expanding homologously.

A broad class of hydrodynamic models involves the formation of bipolar
nebulae by the interaction of a wind with a previously ejected
equatorial density enhancement; essentially, a pre-existing disk or
torus pinches the waist of subsequently ejected material to produce an
hourglass shape.  A model by Morris \& Podsiadlowski (2007) accounts
for the formation of SN~1987A's nebula with a merger of a binary
system, where the merger ejects an equatorial torus or disk of
material that might then divert a faster wind toward high latitudes,
forming a pair of polar caps that could be seen as rings under proper
circumstances of illumination.  That merger model cannot apply here,
since RY Scuti has not yet merged.  More importantly, the gap between
the two components indicates that the second mass ejection is not yet
interacting hydrodynamically with the first one, even at this very
young stage, so the formation of its rings must have a different
origin that doesn't depend on hydrodynamic shaping by a previously
ejected disk.  

Instead, the proper motions in RY Scuti's nebula suggest that on at
least two separate occasions separated by only $\sim 100$--200 yr, the
star system suffered some sort of outburst that ejected mass near the
equator at relatively slow speeds (i.e., much slower than the escape
speed or normal wind speed of either star).  It also suggests that
whatever mechanism shaped the first ejection was able to persist and
have the same influence on the second, because both ejections have the
same basic geometry and structure.  The cause of such an outburst is
unknown, but some ideas are discussed in \S 5.3.  Here we focus on the
shape of the ejecta, and we discuss two potential scenarios, the
second of which we deem to be more likely.

\begin{figure*}\begin{center}
\includegraphics[width=4.3in]{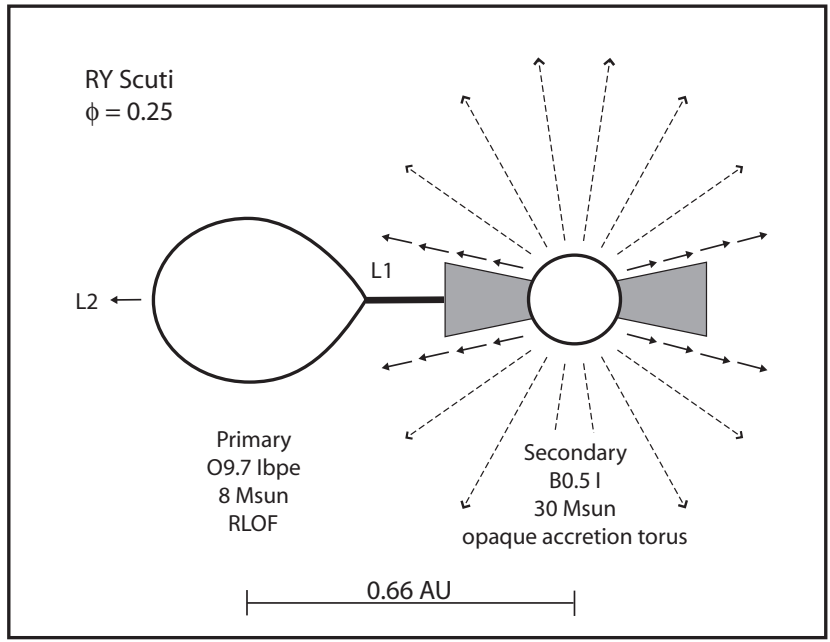}
\end{center}
\caption{Sketch of the RY Scuti system as viewed from the equatorial
  plane near phase $\phi = 0.25$, adapted from the pole-on view of
  Grundstrom et al.\ (2007). The primary is filling its Roche lobe and
  transferring matter to the secondary, which is surrounded by an opaque
  accretion disk/torus.  The long dashed arrows represent the
  trajectory of unimpeded wind or eruptive ejecta from the secondary.
  The shorter solid black arrows represent the trajectory of denser
  and slower disk material entrained by the secondary wind, or
  wind/ejecta that are decelerated through interaction with the torus.
  In a scenario where the accreting secondary suffers an outburst,
  this slower and denser material skimming the edge of the torus may
  form the rings around RY Scuti.  With $M_1 + M_2 = 38 {\rm M}_{\odot}$
  (Grundstrom et al. 2007) and $P = 11.12475$~d (Smith et al.\ 2002),
  the orbital semimajor axis is $a = 0.33$ AU.}
\label{fig:sketch}
\end{figure*}

(1) During an episode of increased mass loss or mass transfer from the
primary star, RY Scuti may have suffered an enhancement of mass loss
through the outer L2 Lagrangian point (see Figure~\ref{fig:sketch}).
From the phase variability of absorption features such as He~{\sc i}
lines, Grundstrom et al. (2007) inferred that RY Scuti does in fact
have a mass-loss stream exiting the system through the L2 point in its
present-day state.  An increase in this mass-loss rate at two separate
times in the past could lead to the creation of discrete toroidal
structures that would expand outward to form the currently observed
circumstellar nebula.  One potential inconsistency is that the
observed outflow through L2 has an expansion speed of
$\sim 200$ km s$^{-1}$ (Grundstrom et al.\ 2007), whereas the toroidal
circumstellar nebula has a much slower radial expansion of only $\sim 40$
km s$^{-1}$ (Smith et al.\ 2002).

Another problem with this scenario is that even with an enhancement of
mass transfer and mass loss through L2, one would normally expect mass
loss through the outer Lagrangian point to be a relatively slow
process compared to the orbital period of only $\sim$11 days, leading
to azimuthally symmetric mass loss.  Thus, while this scenario
accounts for equatorially enhanced mass loss, it provides no
compelling explanation for the nature of sudden outbursts or for the
azimuthal asymmetry observed in RY Scuti's ionised rings.
Furthermore, the origin of the double ionised ring structure is
unclear in this scenario, unless the apparent separation between the
rings is due to a shadow cast by an opaque circumbinary disk at inner
radii much smaller than the size of the rings, as suggested by
Grundstrom et al.\ (2007).

(2) A different scenario may be that the accreting secondary star
experiences an outburst, and that the outflowing ejecta are
immediately shaped by the accretion torus within a few stellar radii
around the secondary (Figure~\ref{fig:sketch}).  As we explain further
in \S 5.3, invoking an outburst from the secondary is motivated by our
speculation that the secondary may encounter a cyclical instability
associated with the high mass accretion and angular momentum accretion
rates currently imposed on it during RLOF.

Suppose that the envelope of the accreting secondary star becomes
unstable and suffers a sudden outburst of mass loss.  The ejecta that
follow mid- to high-latitude trajectories toward the pole will expand
unimpeded (represented by the dashed arrows in
Figure~\ref{fig:sketch}), probably at very high speeds of $\sim 1000$
km s$^{-1}$, close to the escape speed of a main-sequence O-type star
(recall that the secondary star in the RY~Scuti system is thought to be
a main-sequence O-type star that is hidden by its cooler, opaque
accretion torus; Grundstrom et al.\ 2007).  On the other hand, the
ejecta expanding at low latitudes near the equator must contend with
the presence of the massive, dense accretion torus.  The stellar
ejecta will be decelerated by this interaction, and perhaps some of
the material on the surface of the accretion torus will be entrained
by the outflow that is diverted above and below the torus.  This will
result in a much slower outflow from the system at latitudes
immediately above and below the edges of the accretion torus (this is
depicted by the solid short arrows in Figure~\ref{fig:sketch}).  If
the ejection is a sudden event, it will produce an enhancement of mass
at specific latitudes above and below the equatorial plane, at a
specific distance from the star (i.e., plane-parallel rings rather
than a shell or conical structures).\footnote{Note that if the mass
  ejection is related to critical rotation, as we speculate in \S 5.3,
  then it is possible that the mass ejection itself will be inherently
  non-spherical.}  In this scenario, it is the thickness of the
accretion torus that sets the latitude and separation of the rings
that we observe in the circumstellar nebula.  The presence of a thick
accretion torus reaching at least $\pm$15$\arcdeg$ is required by the
fact that the accretion torus obscures the secondary in a system with
an orbital inclination of $i \approx 75\arcdeg$ (Smith et al.\ 2002).
This roughly matches the latitudes of the nebular rings at
$\pm$14$\arcdeg$.

If a sudden (i.e., dynamical) outburst of the secondary star occurs on
a time scale short compared to the orbital period, then it may provide
a compelling explanation for the azimuthal asymmetry in the system as
well: the parts of the outflowing ejecta that expand toward
the bloated supergiant primary star must interact with that star and
its wind.  This would lead to a gap in the ejecta over a range of
azimuthal angles covering 15--30\% of the orbit.  This is commensurate
with the size of the gap on the near side of the nebula (Smith et al.\
2002).  Without a sudden outburst, one would expect the outflow to be
azimuthally symmetric, contradicting observations.

Because a sudden ejection by the secondary star can, in principle,
account for both the latitudinal and azimuthal distribution of mass in
the nebula around RY Scuti, we find this scenario to be more
compelling than option (1) discussed above.  Of course, our suggestion
should be explored using detailed numerical hydrodynamic
simulations.  Some modelers have explored a wind interacting with a
pre-existing disk or torus (e.g., Frank et al.\ 1995; Martin \& Arnett
1995; Blondin \& Lundqvist 1993).  However, these simulations
generally placed the constricting torus at a large distance from the
ejection, rather than an accretion torus within a few stellar radii,
and they adopted a continuous wind that rams into the torus rather
than a sudden ejection (these simulations -- usually in two dimensions -- 
also do not account for the obstruction of a supergiant companion star, of
course).  The sudden ejection is key for both the ring structure and
the azimuthal asymmetry.  The potential for a sudden ejection by the
secondary to account for the structure in RY Scuti's circumstellar
nebula motivates our speculation in \S 5.3 regarding possible physical
causes of an outburst from the accreting secondary.\footnote{Note that
  the apparent thicknesses of the rings and dust torus, which are
  20--25\% of their respective radii, may simply be due to the sound
  speed multiplied by their ages.  The radial expansion speeds are
  40--50 km s$^{-1}$, which is 4--5 times the sound speed when the
  ejected gas is ionised.  In other words, they are as thin as they
  can be, even for an instantaneous ejection.}

\subsection{Significance in Pre-Supernova Evolution, and Post-RLOF
  Binaries as B[e] Supergiants}

Despite the recent episodes of mass ejection, estimates of the nebular
mass compared to the amount of mass exchanged between the stars
suggests that RLOF has been mostly conservative so far for RY Scuti.
The mass of the inner ionised gas torus is roughly 0.003 M$_{\odot}$
(Smith et al.\ 2002), while the mass of the outer dusty torus (from
the measured dust mass multiplied by an assumed gas:dust mass ratio of
100 adopted by those authors) is at least $1.4 \times 10^{-4}$
M$_{\odot}$ (Gehrz et al.\ 2001).  Thus, the total mass detected in
the two-component toroidal nebula around RY Scuti is only of order
0.003 M$_{\odot}$.  This is admittedly a lower limit to the total mass
lost, since there may be dense clumps of neutral gas not measured in
the tracers of ionised gas.  There could also be more mass in the
equatorial plane at larger distances that is shielded from the star's
radiation by the inner components of the nebula, but this mass has not
been constrained by any previous study. In any case, the nebular mass
ejected in the past $\sim$10$^4$ yr is probably far less than
1~M$_{\odot}$.  With a current stellar-wind mass-loss rate of around
(1--2) $\times 10^{-6}$ M$_{\odot}$ yr$^{-1}$ (typical for the wind of
an O9 supergiant; Repolust et al.\ 2004), the mass lost in a fast
line-driven stellar wind during this time should be comparable to that
in the ionised nebula.

As noted in the introduction, however, modern observed parameters for
RY Scuti favour present-day masses of $\sim$30 M$_{\odot}$ for the
secondary, and $\sim$8 M$_{\odot}$ for the primary (originally the
more massive star).  Since the likely initial masses were of order 25
and 15 M$_{\odot}$, this suggests that about 15 M$_{\odot}$ has
been shifted from the primary to the secondary during the brief
($\sim$10$^4$ yr) RLOF phase, whereas much less than 1 M$_{\odot}$
appears to have been lost from the system during that same time.  From
these rough observational estimates, we conjecture that the mass
transfer in RY Scuti has been largely conservative --- that is, unless
a large amount of nebular material resides in the equatorial plane
outside the dust torus where it may be shielded, and may therefore
remain cold and largely neutral.  Observations at far-IR and submm
wavelengths may help constrain the amount of additional cold material
in the system.

While conservative mass transfer rather than mass loss appears to
dominate the stripping of the donor star's H envelope, the mass
ejected into an equatorial torus may play an important role in angular
momentum loss.  The resulting dusty toroid or ring may have observable
consequences after the short phase of RLOF is complete, as noted
earlier.

In the RY Scuti system, the primary has transferred much of its mass
to the secondary, which as a result, is seen as a disk-enshrouded
O-type supergiant star that will now be overluminous for its initial
mass, and will be rapidly rotating due to the additional angular
momentum of the accreted mass (see below).  Eventually, when the
opaque disk dissipates and thins, the secondary will be much brighter
at visual wavelengths than its hotter WR-like primary that will have
lost its envelope, it will have substantially reduced its mass and
luminosity, and it will then radiate most of its luminosity in the
far-UV.  Thus, when RLOF is finished and the primary finally becomes a
WR-like star depleted of H, the overluminous secondary might outshine
the primary at visual wavelengths.  Due to the presence of recently
ejected circumstellar material in a surrounding nebula, the system
will continue to have bright emission lines, radio emission, and IR
excess from dust.  In many observable respects, the system may
therefore resemble a B[e] supergiant (e.g., Zickgraf et al.\ 1996).

If the primary in RY Scuti is, in fact, destined to die as a Type~Ib/c
SN, then it provides us with a real example of what binary progenitors
of SNe~Ib/c may look like.  Additionally, SNe~IIb are closely related
to SNe~Ib, except that they have a small residual H envelope; RY Scuti
could therefore also die as a SN~IIb if it fails to completely shed
its outer H layers before core collapse.  In that case, it provides us
with a Galactic analog to the progenitor of the Type~IIb
explosion SN~1993J in M81 (Filippenko et al.\ 1994), which was
inferred to be a close binary system of slightly lower initil mass
that experienced an almost identical binary evolutionary path (Maund
et al.\ 2004; see also Aldering et al.\ 1994; Van Dyk et al.\ 2002).

A distant extragalactic observer who witnesses the SN resulting from
the explosion of the primary in RY Scuti might be able to infer, from
appropriate pre-explosion archival data, that there was a blue star
with $MV \approx -6$ mag at the same position before the SN.  This
would not have beeen the SN progenitor itself, but the overluminous
mass-gainer secondary in the close binary system.  The alien observer
could verify this conjecture with late-time observations showing that
the blue star was not destroyed in the SN explosion.  The observer
might infer from single-star evolution models that this star had a
ZAMS mass of $\sim 30$ M$_{\odot}$, when in fact the masses of the
primary and secondary have been substantially altered by RLOF. In this
case, however, it would be a mistake to disregard this surviving
source as a chance coincidence, since the overluminous blue companion
provides an important clue that the primary was stripped of its H
envelope via RLOF in a close binary system.  Indeed, Maund et al.\
(2004) identified a massive blue star that might have been the
surviving mass-gainer companion to the star that exploded as SN~1993J.

With only $\sim 0.003$ M$_{\odot}$ of ejected gas in the immediate
circumstellar environment, the shock interaction with the surrounding
torus will not be strong enough to produce a Type IIn supernova (see
Smith et al.\ 2009).  Thus, the resulting explosion from the primary
in RY Scuti will likely be a Type Ib, or perhaps a Type IIb event like
SN~1993J, depending on whether all or nearly all of the H envelope is
transferred from the primary to the secondary.  However, an
interesting consequence of the toriodal circumstellar nebula around RY
Scuti is that the resulting SN might produce a strong IR echo, due to
the circumstellar dust getting heated by the SN's pulse of UV/optical
luminosity.

\subsection{Outbursts from the Accreting Secondary and Massive RLOF
  Binaries as Optical Transients}

Our study provides empirical evidence that the mass-transfer phase in
massive binaries can in some cases be accompanied by episodic bursts
of mass ejection, rather than just a continuous and steady transfer of
mass.  Here we speculate about the underlying physical cause of the
outbursts, and we speculate about possible observed consequences of
these events.


Previous studies of RY Scuti suggest that the initially more massive
star ($\sim$25 M$_{\odot}$) has shed much of its H envelope, leaving
an 8 M$_{\odot}$ stripped-envelope star.  The secondary, initially the
less massive of the two, has already accreted 10--15 M$_{\odot}$
through an accretion disk, yielding a 30 M$_{\odot}$ star that is
still largely obscured by its surrounding accretion torus.  It is
probable that the accreting secondary in such a system will be
significantly spun up due to mass and angular momentum accretion, and
may therefore be at or near critical rotation (e.g., Struve 1963;
Packet 1981; Langer \& Petrovich 2007; Vanbeveren et al.\ 1998; Langer
et al.\ 2008).  Indeed, estimates suggest that accreting even a few to
10\% of a star's mass via an accretion disk in RLOF is enough to spin
up a star to near the critical rotation limit (Packet 1981; Vanbeveren
et al.\ 1998), and this is one of the leading ideas for the formation
of Be stars in binary systems (e.g., Gies 2007; Dewi 2007).  The
secondary star in RY Scuti has accreted a large fraction of its
current stellar mass (roughly half), suggesting that it must have
already encountered an angular momentum catastrophe where it has
reached critical rotation, perhaps repeatedly on several occasions.
Moreover, the addition of mass and heating of the envelope via
accretion luminosity on a short RLOF timescale of less than 10$^4$ yr
[shorter than the (2--3) $\times 10^4$ yr KH time of the entire star]
may leave the envelope overluminous and out of thermal equilibrium
with the core.  Unfortunately, this is difficult to test directly,
since the secondary in the RY Scuti system is hidden by an opaque
accretion disk in its present state.

While in this rapidly rotating nonequilibrium configuration, the
accreting secondary star may be subject to a quasi-cyclical
instability whose recurrence is set by either the angular momentum
diffusion timescale or the thermal timescale in the star's outer
envelope, coupled to the mass and angular momentum accretion rate.
The situation is reminiscent of rotational and thermal instabilities
discussed for the envelopes of LBVs (Appenzeller 1986; Stothers 2000;
Guzik et al.\ 1999; Davidson 1999; Smith et al.\ 2003).  One example
of these is the so-called Omega limit (Langer 1997, 1998; see also
Glatzel 1998), where critical rotation combined with high luminosity
leads to violent mass ejections from a massive star.  Shedding mass
and angular momentum in a shell ejection could temporarily alleviate
the state of critical rotation --- but if the angular momentum of the
secondary is continually replenished with an accretion disk in RLOF,
that star could repeatedly be driven to critical rotation and may
therefore encounter recurring shell ejections.  We suspect that this
scenario might lead to repeated mass ejections like those experienced
by RY Scuti.
With a very high mass-transfer rate of order 10$^{-3}$ M$_{\odot}$
yr$^{-1}$ (i.e., 10--15 M$_{\odot}$ during the RLOF phase of $\sim
10^4$ yr), the secondary of RY Scuti will accrete about 0.1
M$_{\odot}$ over the observed time interval of $\sim$120 yr between
mass ejections.  This mass gained with high specific angular momentum
at the equator is more than the amount of mass lost during the same
time interval, suggesting that it is enough to replenish the amount of
angular momentum that was lost, and would therefore be sufficient to
drive the star back to critical rotation.
Further work on thermal and dynamical instabilities in the envelopes
of rapidly rotating accreting secondaries in binaries would be of
considerable interest.  In particular, a detailed dynamical treatment
of the rotating stellar envelope is needed to constrain the physics of
the instability and whether it can lead to a sudden outburst.  If so,
it might provide a possible explanation for the origin of LBV
eruptions that occur in binary systems.

Recall that our suggestion is motivated in part by the fact that
invoking repeated sudden outbursts from the accreting secondary star
provides a reasonable explanation for several observed properties of
RY Scuti's nebula, including its double-ring toroidal nebula, its
azimuthal asymmetry, and the fact that its repeated mass ejections
that occurred $\sim$250 and $\sim$130 yr ago both had similar
geometry.  The slower leaking of mass through the outer L2 point
provides for equatorially enhanced mass loss, but does not seem to
account for other observed properties of the system (see \S 5.1).  In
the case of RY Scuti, the amount of mass lost appears to be much
smaller than the amount of mass transferred from the primary to the
secondary (i.e., recent RLOF appears to be nearly conservative in the
case of RY Scuti).


Sudden mass ejections are often accompanied by luminous outbursts akin
to the giant eruptions observed in LBVs.  There is a wide diversity of
LBV-like eruptions, sometimes called ``SN impostors,'' which has been
reviewed recently by Smith et al.\ (2011).  The famous massive
binaries $\eta$~Car and HD~5980 both suffered LBV giant eruption
events.  There are a number of other events for which we do not know
whether the progenitors are in binary systems, but two recent examples,
SN~2000ch and SN~2009ip, have exhibited {\it repeating} LBV-like
outbursts (Pastorello et al.\ 2010; Smith et al.\ 2011).  Although
the underlying cause of eruptions is not known, $\eta$ Car exhibited
brief luminous peaks in the light curve at times of periastron in the
eccentric binary (Smith \& Frew 2011; Smith 2011).  These reached
absolute magnitudes of roughly $-$14 mag and lasted for about 100
days, very similar to several other SN impostors.  The relatively
small (0.003 M$_{\odot}$) mass ejections of RY Scuti are not known to
have coincided with a major brightening event, but brief outbursts in
the 18th and 19th centuries might easily have been missed if the
brightening lasted only a few days.  Other more extreme mass ejections
in binaries are possible and do coincide with major brightening
events.

We therefore speculate that some events in the population of observed
extragalactic SN impostors could be related to mass ejections that are
caused by an instability associated with mass and angular momentum
accretion, similar to the one we outlined above for RY Scuti.  Whether
this is true requires more detailed theoretical study of the
dynamical and thermal stability of accreting stars in RLOF.  This
hypothesis has the advantage that accretion-induced critical rotation
can be reached for the mass gainers over a wide range in initial
masses, not limited to the most massive stars near the Eddington
limit.  Many of the SN impostors and related transients appear to
arise from stellar systems with initial masses below 20 M$_{\odot}$
(see Smith et al.\ 2011 and references therein), and this has
been difficult to understand in the context of LBV eruptions.  RY
Scuti presents a concrete example of a RLOF system that has
experienced repeating sudden mass ejections where the mass gainer is
known to be surrounded by an accretion torus.

\section{CONCLUSIONS}

We briefly summarise the main observational conclusions and their
implications from our study of the expansion of RY Scuti's nebula.

(1) The expansion age of the inner ionised rings measured using {\it
  HST} images indicates an ejection date of roughly 1881 ($\pm$4 yr).

(2) The expansion age of the outer dusty IR torus measured using Keck
AO images yields a likely ejection year of 1754 ($\pm$36 yr).

(3)  We therefore conclude that the two components of RY Scuti's nebula
(the inner ionised rings and the outer dust torus) were
the result of two separate ejection events recurring on a timescale of
$\sim 120$--130 yr.  One may wonder whether RY Scuti is due for another such
ejection event in the near future.

(4) Conclusion (3) is supported by a clear spatial gap between the two
structural components, indicating that they are not interacting
hydrodynamically.  Therefore, one cannot invoke a distant pre-existing
disk as the shaping mechanism for the double-ring nebula.  This gap
also indicates that few ionising photons can penetrate the inner
ionised rings.  We speculate that the second ejection event, which is
now seen as the expanding ionised rings in {\it HST} images, may have
shielded the outer torus from the stars' ionising radiation, thereby
allowing dust to form in the outer torus.

(5) We suggest a formation mechanism for the toroidal circumstellar
nebula that involves matter ejected suddenly (i.e., dynamically) by the
accreting secondary star, and immediately being shaped by the opaque
accretion torus that is thought to surround the secondary star,
although we encourage numerical simulations of this scenario.

(6)  The primary star in RY Scuti, which is in the process of losing
its H envelope via RLOF, is our best-studied candidate for the progenitor 
of a Type IIb or Type~Ibc supernova where the envelope stripping from
close binary evolution is currently underway.  While other post-RLOF
systems are good candidates for SNe~Ibc as well, RY Scuti is a
rare example of a system that is currently in the critical mass
transfer phase, and which is surrounded by a nebula that allows us to
measure the mass lost from the system.

(7)  RY Scuti suggests, therefore, that the formation of SN~Ibc
progenitors via RLOF may in some cases be punctuated by sudden,
repeating episodes of mass loss.  We speculate that this may be the
result of an instability associated with mass and angular momentum
accretion by the secondary star, although this idea deserves
additional study.  In some cases these mass-ejection events may be
seen as luminous outbursts, even though we are aware of no such record
of an observed outburst in the specific case of RY Scuti.  It is worth
considering the possibility that other close binary systems may
contribute to the diverse population of non-supernova optical
transients now being discovered.

(8) When RLOF finishes for RY Scuti, we speculate that the accretion
torus around the secondary will become optically thin, revealing the
rapidly rotating and overluminous mass gainer.  The system will also
be surrounded by a dusty torus resembling those seen in B[e] stars. If
this overluminous 30 M$_{\odot}$ secondary star dominates the optical
luminosity of the system, it may point toward an observed association
between OB emission-line stars and some SN~Ibc progenitors, even
though it is the companion to the overluminous OB star that actually
explodes.

\smallskip\smallskip\smallskip\smallskip
\noindent {\bf ACKNOWLEDGMENTS}
\smallskip
\footnotesize

Support was provided by the National Aeronautics and Space
Administration (NASA) through grants GO-11977, GO-8209, and GO-6492
from the Space Telescope Science Institute, which is operated by AURA,
Inc., under NASA contract NAS5-26555.  RDG was supported by NASA and
the United States Air Force.  Some of the data presented herein were
obtained at the W.M.\ Keck Observatory, which is operated as a
scientific partnership among the California Institute of Technology,
the University of California, and NASA.  The Observatory was made
possible by the generous financial support of the W.M. Keck
Foundation.  The authors wish to recognise and acknowledge the very
significant cultural role and reverence that the summit of Mauna Kea
has always had within the indigenous Hawaiian community; we are most
fortunate to have the opportunity to conduct observations from this
mountain.


\end{document}